\documentclass[prl,twocolumn]{revtex4}

\usepackage[dvips]{graphicx}%
\usepackage{bm,color}
\usepackage{amsmath,amssymb}

\newcommand{\ketbra}[2]{\ket{#1}\!\bra{#2}} %\newcommand{\ketbra}[3]{\ket{#1}_{#3}\bra{#2}}
\newcommand{\ket}[1]{\left |  #1 \right \rangle}
\newcommand{\bra}[1]{ \left \langle #1  \right |}
\newcommand{\ave}[1]{ \langle #1  \rangle}
\def \tr{{\textrm {Tr}}}

\begin{document}
%%%%%%%%
%  \small
%%%%%%%%
\title{Bayesian uncertainty relation for a joint measurement of canonical variables}

\author{Ryo Namiki}%
\affiliation{Department of Physics, Gakushuin University, 1-5-1 Mejiro, Toshima-ku, Tokyo 171-8588, Japan}
%\email[Electric address: ]{namiki@qi.mp.es.osaka-u.ac.jp}%
 %\email[Electric address: ]{namiki@qi.mp.es.osaka-u.ac.jp}%

\date{\today}%{September 13, 2010}%\today}
\begin{abstract} 
We present a joint-measurement uncertainty relation for a pair of mean square deviations of canonical variables averaged over Gaussian distributed quantum optical states. 
 Our Bayesian formulation is free from the unbiasedness assumption,  and enables us to quantify  experimentally implemented joint-measurement devices by feeding a moderate set of coherent states. Our result also reproduces the most informative bound for quantum estimation of phase-space displacement in the case of pure Gaussian states. 
\end{abstract}

% \pacs{03.67.Dd, 42.50.Lc} 
%\keywords{quantum cryptogrphy}
\maketitle

%\newpage
%%%%%%%%%%%%%%%%%%%%%   part 1 %%%%%%%%%%%%%%%%%%%%%%%%%%%%%%%%%%%%%%%%%%%%%%%%
%%%%%%%%%%%%%%%%%%

%\small 
%\subsubsection{Introduction}
%Abstract: \textit{}

The canonical commutation relation $[\hat x, \hat p]= i \hbar$ determines a primary noise property on quantum states, and it is highly fascinating if one can characterize the penalty on controlling or measuring a pair of canonical variables  in a unified manner \cite{Busch14,AG88,Cle10}. The amplification uncertainty relation represents the penalty of the linear amplification by a gain dependent uncertainty relation \cite{Amp,Cle10}%{
\begin{align}
\ave{(\hat x - \sqrt{G_x }x_{\textrm{in}})^2}\ave{(\hat p - \sqrt{G_p } p_{\textrm{in}})^2} \ge  \frac{\hbar^2}{4} \left( { G +|G-1 | }   \right) ^2,   \label{AUR0}
\end{align}
%\begin{align} \ave{(\hat x - \sqrt{\tau_x }x_{\textrm{in}})^2}\ave{(\hat p - \sqrt{\tau_p } p_{\textrm{in}})^2} \ge  \hbar^2 \left(\frac{ \tau +|\tau-1 | }{2 }  \right) ^2,   \label{AUR0}\end{align}
 where  $(G_x,G_p )$ represents the gain of the physical process with $G=\sqrt{G_x G_p}$, and $(x_{\textrm{in}},p_{\textrm{in}})$ stands for the mean value of  $(\hat x , \hat p)$ for an arbitrary input state. The amplification map is assumed to satisfy the  covariance condition, $\ave{\hat x}= \sqrt {G_x}x_{\textrm{in}}$ and $\ave{\hat p}= \sqrt {G_p } p_{\textrm{in}}$.   
Under the  covariance condition, a class of quantum channels composed of a state-preparation process after a measurement process referred to as entanglement breaking channels is suggested to  fulfill %an analogous inequality
 \cite{Braunstein-Kimble98,Hol08,Namiki-Azuma13x} %,Namiki-Azuma13x}
\begin{align} \ave{(\hat x - \sqrt{G_x }x_{\textrm{in}})^2}\ave{(\hat p - \sqrt{G_p } p_{\textrm{in}})^2} \ge  \frac{\hbar^2}{4} \left( 2G +1 \right) ^2. \label{AUR1} \end{align}
%\begin{align} \ave{(\hat x - \sqrt{\tau_x }x_{\textrm{in}})^2}\ave{(\hat p - \sqrt{\tau_p } p_{\textrm{in}})^2} \ge  \hbar^2 \left(  \tau + \frac{1 }{2 } \right) ^2. \label{AUR1} \end{align}
The right-hand value of Eq.~\eqref{AUR1} is two units of the shot noise  $2 \times (\hbar/2) $ larger than the right-hand side of Eq.~\eqref{AUR0} in the case of a unit gain $G=1 $  \cite{Braunstein-Kimble98,Hol08,Namiki-Azuma13x}. For a measurement process, the famous joint-measurement uncertainty relation \cite{AK65} can be expressed as 
\begin{align}
\ave{(\hat Q_1 -x_{\textrm{in}})^2}\ave{(\hat P_2 -p_{\textrm{in}})^2} \ge  \hbar ^2, \label{measUR}  %\left(\frac{\hbar }{2}
\end{align} 
where the variables of two meter systems $\hat Q_1$ and $\hat P_2$  are assumed to satisfy the unbiasedness condition, $\ave{\hat Q_1}=x_{\textrm{in}}$ and  $\ave{\hat P_2}=p_{\textrm{in}}$. % where  $\hat Q_1$ and $\hat P_2$ are observables of two meter systems. 
The right-hand term of Eq.~\eqref{measUR} becomes larger-than the right-hand term of Eq.~\eqref{AUR0} with $G =1 $ by a single  unit of the shot noise $\hbar/2$. 
As a standard interpretation, this extra noise originates from an introduction of the auxiliary system, and another extra unit for entanglement breaking channels in Eq.~\eqref{AUR1} is due to the state-preparation process. %

Unfortunately,  neither the covariance  condition nor the unbiasedness condition for continuous-variable (CV) systems is realized in experiments since such a condition implies %input states of unbounded energy and 
devices executing a perfectly linear response to arbitrary field amplitude.  
Hence, there is no reason to consider that the inequalities in Eqs.~\eqref{AUR0},~\eqref{AUR1}, and~\eqref{measUR} hold for real physical systems. In addition, the amplitude of input states is practically bounded due to   physical conditions in experimental systems, and it is impossible to confirm the linearity. This motivate us to search for theoretical limitations on physical maps assuming a moderate set of input states \cite{Appleby98a}. For optimization of the joint measurement, a Bayesian approach for input-state ensembles of coherent states with Gaussian prior had been studied in depth in the seminal results of quantum estimation \cite{holevo1973optimal,Yuen-Lax,holevo1975some,holevo2011probabilistic,Helstrom}. Due to the Bayesian framework, the measurement devices need not to be unbiased, and the Gaussian prior enable us to avoid the contribution from unrealistically higher input amplitude. Such Bayesian state ensembles have also been employed to determine classical limits or fundamental quantum limits for continuous-variable (CV) quantum gates mainly in terms of the average gate fidelity \cite{Ham05,Namiki07,Owari08,Namiki11R,Namiki11,Namiki2011,Chir13,Chiribella2014a,
Yang2014,Namiki1511,ZhaoChiribella17,Namiki1503}. Recently, the uncertainty-relation type inequalities essentially the same form as Eqs.~\eqref{AUR0} and~\eqref{AUR1} have been proven in terms of the mean-square deviations (MSD)  for canonical variables
in Refs.~\cite{Namiki1502} and  \cite{Namiki-Azuma13x}, respectively.

In this report, we derive a joint-measurement uncertainty relation for Bayesian MSDs %mean-square deviations (MSDs),
 which coincides with the form of the Arthurs-Kelly relation in Eq.~\eqref{measUR}. %
It turns out that the three uncertainty relations in  Eqs.~\eqref{AUR0},~\eqref{AUR1}, and~\eqref{measUR} can be reformulated to be applicable for a wide range of physical processes realized in experiments.
 The inequality is proven as a straightforward consequence of commutation relations and the property of quantum channels. Further, our measurement model is proven to be equivalent to the model of parameter estimation based on the positive operator valued measure (POVM).
Our result also reproduces the most informative bound for quantum estimation of phase-space displacement in the case of pure Gaussian states \cite{holevo2011probabilistic}.

In our approach \cite{Namiki1502, Namiki-Azuma13x}, we consider the MSDs for a set of coherent states instead of the deviations for an arbitrary state in the left-hand side of Eqs.~\eqref{AUR0},~\eqref{AUR1}~and~\eqref{measUR}. In what follows we set $\hbar  =1 $ and the  canonical  commutation relation indicates  $[\hat x, \hat p ]= i $  as a standard notation.
The mean input quadrature $(x_{\textrm{in}},p_{\textrm{in}})$ for a coherent state $ \rho_\alpha := \ketbra {\alpha}{\alpha}$ is specified by 
\begin{eqnarray}
x_\alpha  := \tr (\hat x \rho_\alpha ) %\ave{\hat x }_{\rho_\alpha} 
= \frac{\alpha + \alpha ^*}{\sqrt 2 }, \ p_\alpha := \tr (\hat p \rho_\alpha ) %\ave{\hat p }_{\rho_\alpha} 
= \frac{\alpha - \alpha ^*}{\sqrt 2 i }.  \label{ShNote}  % \nonumber \\ \rho _\alpha &=&  \ketbra{\alpha }{\alpha }, \ket{\alpha} = e^{-|\alpha |^2/2} \sum_{n=0}^\infty.  \frac{\alpha ^n }{\sqrt {n! }}\ket{n}
\end{eqnarray} % represent.  Here and 

Let $\hat M $ and $\hat N$ be self-adjoint operators, and $\mathcal E$ be a quantum channel.  We define a pair of the MSDs as 
\begin{eqnarray}
 \bar V_M^{(x)} (\eta_x, \lambda) &:=& \tr \int  p_\lambda (\alpha )   (\hat M  -   \sqrt \eta_x  x_\alpha )^2   \mathcal E ( \rho _\alpha ) d^2 \alpha  \nonumber \\ 
  \bar V_N^{(p)} (\eta_p, \lambda) &:=  &  \tr \int  p_\lambda (\alpha )   (\hat N  -   \sqrt \eta_p p_\alpha )^2   \mathcal E ( \rho _\alpha ) d^2 \alpha,    \label{defbarxp} \end{eqnarray}
where  %the gain 
 $(\eta_x, \eta_p)$ is  a pair of non-negative number, and % is given by  
\begin{eqnarray}
p_\lambda( \alpha ) :=  \frac{\lambda }{\pi} \exp (- \lambda |\alpha |^2 ) \label{eq2}\end{eqnarray} is  a Gaussian prior distribution with %an inverse width 
$\lambda >0 $. 
If we choose the canonical pair $(\hat M, \hat N) =(\hat x, \pm \hat p)$, % and $\mathcal E$ is a CPTP map, 
we can reach the Bayesian amplification uncertainty relation for quantum channels \cite{Namiki1502} 
\begin{eqnarray}
 \bar V_x^{(x)}  (G_x, \lambda ) \bar V_{\pm p}^{(p)}  (G_p, \lambda )\ge \frac{1}{4}  \left( \frac{  G }{1+ \lambda }+  \left|\frac{ G }{1 + \lambda }  \mp 1 \right|  \right)^2, \label{B1}
 \end{eqnarray} where the lower sign corresponds to the case of phase-conjugate amplification and attenuation. Moreover, if $\mathcal E$ is entanglement breaking, the minimum of the product $\bar V_x \bar V_p $ has to satisfy a more restricted condition \cite{Namiki-Azuma13x}: \begin{eqnarray}
 \bar V_x^{(x)} (G_x, \lambda ) \bar V_p^{(p)} (G_p, \lambda )\ge \frac{1}{4} \left( \frac{2 G }{1+ \lambda }+1 \right) ^2.  \label{B2} %\frac{\ 3^2}{4}. 
 \end{eqnarray} Notably, Eqs.~\eqref{B1} and \eqref{B2} reproduce the forms of Eqs.~\eqref{AUR0}~and~\eqref{AUR1} in the limit of the uniform prior $\lambda \to 0$. 
In what follows, we will find that a general joint measurement, which is described by a quantum channel $\mathcal E$ and commutable observables $ [\hat M, \hat N ] =0$, has to fulfill 
\begin{eqnarray}
 \bar V_M^{(x)} (G_x, \lambda ) \bar V_N^{(p)}  (G_p, \lambda )  \ge  \left(\frac{G}{1+ \lambda }\right)^2 . \label{B3}  %\frac{\ 2^2}{4} =1. % , 
 \end{eqnarray} % where for observed.
This relation  will establish a joint-measurement uncertainty relation free from the unbiasedness assumption. The inequality in Eq.~\eqref{B3} also reproduces the form of Eq.~\eqref{measUR}  in the limit  $\lambda \to 0$ for $G =1$.

We will find a bound on the pair of the MSDs based on the method in Refs.~\cite{Namiki1502,Namiki-Azuma13x}. Let  be $ \ket{\psi_\lambda}= \sqrt{\lambda /(1+ \lambda )}\sum_{n=0}^\infty (1+ \lambda)^{-n/2}  \ket{ n}_A \ket{ n }_B$ a two-mode squeezed state, and consider the  
 state after an action of  the quantum channel $\mathcal E$ on subsystem $A$, 
\begin{align}
J =&  \mathcal E_A  \otimes  I_B  ( \ketbra{\psi_\lambda}{\psi_\lambda} ), % /P_s, \nonumber \\  P_s&:=   \tr[\mathcal E_A  \otimes  I_B  ( \ketbra{\psi_\lambda}{\psi_\lambda} )] =  \int p_\lambda (\alpha ) \mathcal E (\rho_\alpha)  d^2 \alpha, 
\label{ChoiSt}
\end{align} where $I$ denotes the identity map. 
We can observe that the MSD represents the correlation between a  measurement observable on $A$ and a canonical variable on $B$ for the  state $J$. To be concrete, a straightforward calculation \cite{Namiki1502, Namiki-Azuma13x} from Eqs.~\eqref{defbarxp},~\eqref{eq2},~and~\eqref{ChoiSt}, leads to   
%\textbf{Lemma.--- }
\begin{align} \bar V_M^{(x)} %^{({ \rm Prob})}
  (\eta_x, \lambda) =& \tr\left[( \hat M_A - \sqrt{\tau_x }\hat x_B)^2 J \right] +  \frac{\tau_x}{2 }  %  \left[  \bar V_p (G, \lambda) - \frac{G}{2 (1+ \lambda )} \right] 
 ,     \nonumber \\
\bar V_N^{(p)} %^{({ \rm Prob})}
  (\eta_p, \lambda) = &\tr\left[( \hat N_A + \sqrt{\tau_p} \hat p_B)^2 J \right] +  \frac{\tau_p}{2}  %  \left[  \bar V_p (G, \lambda) - \frac{G}{2 (1+ \lambda )} \right] 
 ,     \label{tekito}  \end{align}
where the rightmost terms are responsible for the vacuum fluctuation  due to  the  mapping procedure from the q-numbers to c-numbers  $(\hat x , \hat p) \to (x_\alpha, p_\alpha)$, and  we defined  
% $J =  \mathcal E_A  \otimes  I_B  ( \ketbra{\psi_\xi}{\psi_\xi} ) /P_s $ %where  $ \ket{\psi_\xi}= \sqrt{1-\xi^2 }\sum_{n=0}^\infty \xi ^n  \ket{ n} \ket{ n }$ is a two-mode squeezed state with $\xi \in (0,1)$ and $P_s:=  \tr[\mathcal E_A  \otimes  I_B  ( \ketbra{\psi_\xi}{\psi_\xi} )]$.
\begin{align}
%(  \widetilde \upsilon_M^{(x)},  \widetilde \upsilon_N^{(p)}) : = (  \bar V_M^{(x)},  \bar V_N^{(p)}) / P_s,\nonumber \\
(\tau_x, \tau _p) : = (  \eta_x, \eta_p ) /  (1+ \lambda ). \label{defcon}%^{-1} , \ \xi = (1+ \lambda)^{-1}. 
\end{align}
A physical limitation  in a product form is  directly imposed by using the preparation uncertainty relation for the state $J$: 
\begin{align}\label{usr1}
 &  \tr [ (   \hat M  _A-  \sqrt{\tau_x} \hat x_B )^2 J ] \tr [ (  \hat N _A+   \sqrt{\tau_p} \hat p_B )^2 J ]  \nonumber \\ 
   \ge&    \ave{\Delta ^2 (\hat M_A-  \sqrt{\tau_x} \hat x_B )}_J \ave{\Delta ^2 (\hat N_A+   \sqrt{\tau_p}  \hat p_B )}_J   \nonumber \\
    \ge& \frac{1}{4} \left| \ave{  [\hat M_A, \hat N_A] \otimes \openone_B -  \sqrt{\tau_x \tau _p} \openone_A  \otimes  [\hat x_B, \hat p_B  ] }_J      \right| ^2. %\label{preLemma} %\openone_B 
 \end{align}

%\begin{widetext}
\textbf{Lemma.--- }
Let $\hat M $ and $\hat N$ be a pair of  self-adjoint operators, and  $\mathcal E$ be a quantum channel. For any given  positive numbers $( \lambda,  \eta_x,   \eta_p   )$, the following relation holds: 
\begin{align}\label{AUP2} 
 &\left( \bar V_M^{(x)}(\eta_x,\lambda ) - \frac{\eta_x}{2 (1+ \lambda )}  \right) \left (\bar V_N^{(p)}(\eta_p,\lambda ) -  \frac{\eta_p}{2 (1+ \lambda )}  \right) \nonumber \\ 
 &  %\nonumber \\
\ge \frac{1}{4} \left| \ave{[ \hat M, \hat N]}_{ \tr_B[ J]} %{\mathcal E({\bar \rho_{ \lambda}})} 
 - \frac{  \sqrt{\eta_x \eta_p}   }{(  1+ \lambda  )   } \right | ^2,   \end{align}  where  the MSDs $(  \bar V_M^{(x)} ,  \bar V_N^{(p)})$ %$(  \bar \upsilon_M^{(x)} ,  \bar \upsilon_N^{(p)})$ %$\bar V_z ^{(\textrm{Prob})} = \bar V_z /P_s $
 is defined in %through  %Eqs.~\eqref{defbarxp} and 
 Eq.~\eqref{defbarxp}, and the state $J$ is given in Eq.~\eqref{ChoiSt}. Note that we can readily prove a similar relation when %$P_s =1 $ whenever 
$\mathcal E$ is a stochastic quantum channel (See  \cite{Namiki1502,Namiki-Azuma13x,Namiki1503}).

\textbf{Proof.}---  Substituting Eqs.~\eqref{tekito}~and~\eqref{defcon} into Eq.~\eqref{usr1} with the help of the canonical commutation relation   $ [\hat x_B, \hat p_B ]  =i   $, we obtain the inequality of Eq.~\eqref{AUP2}.
   \hfill$\blacksquare$

When we set $(\hat M, \hat N) = (\hat x, \pm \hat p)$,  Lemma  leads to  the Bayesian amplification uncertainty relation in Eq.~\eqref{B1} \cite{Namiki1502}. Similar setting enables us to derive Eq.~\eqref{B2} where a separable inequality for Einstein-Podolsky-Rosen like operators \cite{Giov03,Namiki13J} is employed 
  instead of Eq.~\eqref{usr1} \cite{Namiki-Azuma13x}.

  Our interest here is to address the joint-measurement uncertainty relation in the form of Eq.~\eqref{B3}. In a general setup of joint measurements,  the signal state in system $A$ is interacted with an ancilla system  $A^\prime$. Then, a projective measurement $\hat Q$ concerning the original, or true position of the signal is performed on the system $A$, and another projection $\hat P$ concerning the momentum of the original signal is carried out on system $A ^\prime $. The measurement observables are typically written as 
  \begin{align}  \hat M= \hat Q_A \otimes \openone _{A^\prime }, \  %$ and $
\hat   N= \openone_{A} \otimes \hat P_{A^\prime} , \label{typicalM}
  \end{align} and thus  commutable $[ \hat M, \hat N] =0  $.

Here, we describe the interaction with possible ancillary systems by a quantum channel $ \mathcal E$, in which an input state in a single mode system could  be transformed into an output state in any physically allowable system of an arbitrary size. As for the measurement observables we only assume that they are commutable $[ \hat M, \hat N] =0 $.

\textbf{Proposition.---} Let $(\hat M , \hat N) $ be a pair of commutable observables satisfying $[\hat M, \hat N] =0$ and $\mathcal E$ be a quantum channel. For any given positive numbers  $( \lambda,  \eta_x,   \eta_p   )$,
 the following trade-off  relation holds %{\small 
\begin{align}  &\left(\bar V_M^{(x)}  (\eta s , \lambda)- \frac{1}{2}\frac{\eta s}{1\! +\!  \lambda }\right) \!\! \nonumber \\  & \times  \left(\bar V_N^{(p)}  (\eta / s  , \lambda)- \frac{1}{2}\frac{\eta /s }{1\! +\!  \lambda }\right)  
%\bar V_N  (\eta e^{ 2R} , \lambda)    %  \left[  \bar V_p (G, \lambda) - \frac{G}{2 (1+ \lambda )} \right] 
% \nonumber \\ 
\ge  \frac{1}{4} \left ( \frac{  \eta  }{   1\! +\!  \lambda   }\right ) ^2\! ,    \label{SUR2}  \end{align} %}
where   the MSDs $(  \bar V_M^{(x)} ,  \bar V_N^{(p)})$ %$\bar V_z ^{(\textrm{Prob})} = \bar V_z /P_s $
 are defined in   %Eqs.~\eqref{defbarxp} and 
 Eq.~\eqref{defbarxp}. 
 Moreover, the equality of Eq.~\eqref{SUR2}  can be achieved by a joint-measurement setup  using a beam splitter and quadrature measurements. 

\textbf{Proof of Proposition.---}Substituting $ (\eta _x, \eta_p ) = \eta ( s, s^{-1})  $ and  %the commutable condition
 $[ \hat M, \hat  N ] = 0$ into our Lemma of Eq.~\eqref{AUP2}, we obtain Eq.~\eqref{SUR2}.  
In order to prove the attainability, let us consider a half-beam splitter with a vacuum field as an ancilla that transforms the coherent state  as   \begin{align}
\mathcal E (\rho _ \alpha ) = \rho _ {\alpha / \sqrt 2 } \otimes \rho _ {\alpha / \sqrt 2 }. \label{ach1} \end{align} Let us set  the pair of observables as  
\begin{align}
(\hat M,\hat N) &= \frac{\sqrt{2 \eta }}{1+ \lambda}  ( s^{1/2} \hat x     \otimes  \openone , s^{-1/2} \openone \otimes   \hat p) . %,\nonumber \\  
%\hat N &=  g e ^R \openone \otimes   \hat p. 
   \label{ach2}    
\end{align}  It clearly fulfills $[\hat M, \hat N ] =0$. Substituting Eqs.~\eqref{ach1}~and~\eqref{ach2} into Eq.~\eqref{defbarxp}, we obtain 
\begin{align}
  \left(\bar V_M  (\eta s , \lambda), \bar V_N  (\eta / s  , \lambda) \right) &= \frac{ {  \eta }}{1+ \lambda}  ( s , s^{-1} ) . %,\nonumber \\  
%\hat N &=  g e ^R \openone \otimes   \hat p. 
   \label{ach2-}    
\end{align} 
This saturates   the inequality in  Eq.~\eqref{SUR2}.   
Note that the ratio $s= \sqrt{\eta_x / \eta_p}$ can be interpreted as a consequence that either the state is squeezed or the measured value is rescaled.  
\hfill$\blacksquare$

\begin{figure}[tbph]
  \begin{center}
\includegraphics[width=0.8\linewidth]{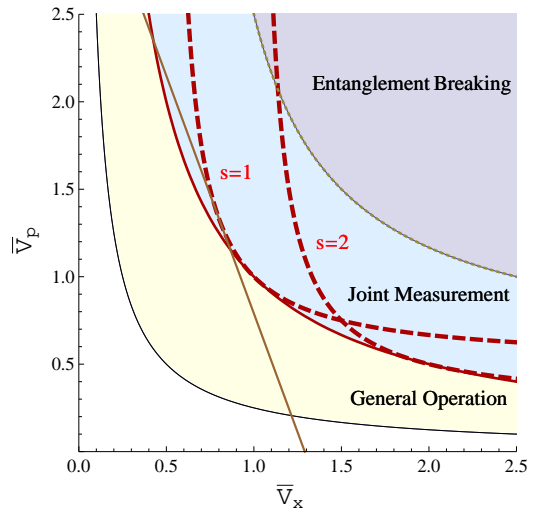}%{smfig1.eps}%{ampfig1.eps}
  \end{center}
  \caption{Interrelation between the minimum uncertainty  curves for quantum channels, joint-measurements, and entanglement-breaking maps in the case of unit gain and uniform prior $(G, \lambda)=(1,0)$
 in  Eqs.~\eqref{B1}, \eqref{B2}, and \eqref{B3}.
The curve for the joint-measurement process is obtained by tweaking $s \in(0,\infty)$ in Eq.~\eqref{SUR2}. Here, the cases of  $ s=1$ and $s=2$ are displayed as the dashed curves. Tangent lines of these curves correspond to the bound for the weighted mean-square errors in the multi-parameter estimation. 
 }
    \label{fig:smfig1.eps}
\end{figure}

Proposition implies that the product $\bar V_M   \bar V_N  $ is lower bounded as in Eq.~\eqref{B3} when the gain $\eta = \sqrt{\eta_x \eta_p} $ is fixed and the ratio $s= \sqrt{\eta_x / \eta_p}$ is tweaked  as in Fig.~\ref{fig:smfig1.eps} (See also Sec.~IIB of \cite{Namiki1502}). Moreover, the inverse proportional curve due to Eq.~\eqref{B3} can be swept by Eq.~\eqref{ach2-}. Therefore, this completes the final step to reformulate the three uncertainty relations of Eqs.~\eqref{AUR0},~\eqref{AUR1}, and \eqref{measUR} into experimentally testable relations of Eqs.~\eqref{B1},~\eqref{B2}, and~\eqref{B3} based on a unified framework without imposing the linearity assumptions on the physical maps. We can observe that the minimum penalty curve for the measurement process is located in the middle of the curves for the quantum channel and the entanglement breaking channel (See Fig.~\ref{fig:smfig1.eps}). This is a rigorous example that shows a hierarchy on the trade-off relations of controlling canonical variables for general quantum channels, joint measurements, and entanglement breaking channels suggested from the uncertainty relations in Eqs.~\eqref{AUR0},~\eqref{AUR1},~and~\eqref{measUR}.

 %%%%%%%%%%%%%%%%%%%%%%%%%%%%%
%%%%%%%%%%%%%%%%%%%%%%%%%%%%%
%%%%%%%%%%%%%%%%%%%%%%%%%%%%%
%%%%%%%%%%%%%%%%%%%%%%%%%%%%%

 Thus far, we have described the measurement process by using a  quantum channel $\mathcal E$ and a pair of observables, $\{\mathcal{E},\hat M, \hat N \}$. In the following part, we rewrite our measurement model by using the POVM. 
This will establish a link between our measurement uncertainty relation and a major result of the quantum estimation theory  \cite{holevo1975some}. Let us start with recalling the framework of parameter estimation.  We consider a set of quantum states called parametric family $\{ \rho_\theta \}_{\theta \in \Theta}$, where ${\theta}$ is an unknown parameter belongs to a set $ \Theta$. A prior probability distribution  $\{p_\theta\}_{\theta \in \Theta}$ is assigned to specify an ensemble of states $\{p_\theta,  \rho_\theta  \}_{\theta \in \Theta}$ in Bayesian approach. The true value of an operator $\hat X$ for an unknown state $\rho_\theta $ is defined as $X_\theta = \tr (\rho_\theta \hat X )$.  An estimation process is described by a POVM $\{\hat m_i\}$ and  a set of real quantities $\{X_i\}$; The estimator $\{\hat m_i,X_i\}$ %assigns process infers the true value  $X_\theta$ by assigning  
an estimation value $X_i$ for each measurement outcome $i$ of the POVM. 
  An estimator is said to be optimal if it %A main goal of parameter estimation is to
 minimizes  the mean square error (MSE) \begin{align}
V_X:= \sum_{\theta\in\Theta  } \sum_i p_\theta (X_\theta - X_i)^2 \tr(\hat m_i \rho_\theta), \label{QESmse}
\end{align} where both the set $\{X_i\}$ and the POVM $\{\hat m_i\}$ are  optimized. Roughly speaking, %we would like to find a better POVM that infers the true value $X_\theta$ over the source of incoming state statistically described by  $\{p_\theta,  \rho_\theta  \}_{\theta \in \Theta}$.
 the estimator $\{\hat m_i, X_i\}$ is more favorable if  the loss function $V_X$ is smaller. 
For the multi-parameter estimation, we may consider a set of operators $\{ \hat X, \hat Y, \hat Z, \cdots \} $, and the problem is to  minimize a wighted sum of the mean square errors for a set of operators such as $g_X V_X+ g_Y V_Y+g_Z V_Z +...$, where $g_{X,Y,Z, ...} \ge 0$ denotes the wighting factors, and a multi-parameter estimator can be specified by the sequence $\{\hat m_i, X_i,Y_i,Z_i, ...\}$.

Our Proposition leads to a bound for a product of the MSEs for the canonical observables $(\hat x, \hat p)$.

\textbf{Corollary.--- }
Let be $\lambda > 0$, $G> 0$, and $R \in \mathbb R$.
Let  $\{ \hat m_i \}$ be a POVM.  For any multi-parameter estimator described by $\{\hat m_i, X_i, P_i\}$, the following trade-off relation  holds   
   \begin{align}
V_X V_P \ge \left ( \frac{ G  } {1+ \lambda} \right )^2 \label{cororor}
\end{align} 
where the pair of the  MSEs are given by %for $\{p_\lambda(\alpha), \rho_\alpha \}_\alpha$ 
  \begin{align}
V_X:= &  \sum_i \int  p_\lambda (\alpha )  (  X_i     -  \sqrt G x_ \alpha e^{-R} )^2  \tr( \hat m_i   \rho _\alpha )   d^2 \alpha , \nonumber  \\
V_P:=&   \sum_i  \int  p_\lambda (\alpha )  (  P_i     - \sqrt G p_ \alpha  e^{ R})^2  \tr( \hat m_i   \rho _\alpha )   d^2 \alpha   .  \label{VXP}
\end{align}
Note that one can drop the parameters $(G, R)$ by considering a rescaled estimation with $(\tilde X_i, \tilde P_i )= G^{-1/2}(X_i e^{R}, P_i e^{-R})$. In such a scenario we have $\tilde V_X  \tilde V_P \ge  ({1+ \lambda})^{-2}$ with $(\tilde V_X, \tilde V_P )= G^{-1}(V_X e^{2R}, V_P e^{-2R})$, instead of Eq.~\eqref{cororor}. This is indeed in  the form of Eq.~\eqref{measUR}. 

\textbf{Proof.---}
Let us define an entanglement breaking channel associated with the POVM $\{ \hat m_i \} $ as 
   \begin{align}
\mathcal E  (\rho ) = \sum_i \tr (\hat m_i \rho ) \ketbra{u_i}{u_i } \otimes  \ketbra{v_i}{v_i }, \label{PC1}
\end{align}  where $\{ \ket{u_i}\} _i $ and $\{ \ket{v_i} \} _i$ are orthonormal bases.  Let us choose a pair of commutable observables as 
\begin{eqnarray}
\hat M%&
=   \sum_i  (X_i  \ketbra{u_i}{u_i }  )  \otimes  \openone  , \ % \ \nonumber \\  
\hat N %&
= \openone \otimes  \sum_i P_i  \ketbra{v_i}{v_i }  \label{PC2} .  
\end{eqnarray} % Obviously, $\hat M$ and $\hat N$ are commutable. 
By substituting Eqs.~\eqref {PC1} and \eqref{PC2}  into Eq.~\eqref{defbarxp} with   $(\eta_x ,\eta_p ) = (G e ^{-2 R} ,G e ^{2 R} )$, %  $\bar V_M  (\eta e ^{-2 R} , \lambda)$  and $ \bar V_N  (\eta e^{ 2R} , \lambda)$ 
 we obtain $V_X$ and $V_P$ in Eqs.~\eqref{VXP}, i.e., we can write  $(V_X, X_P) = (\bar V_M  (G e ^{-2 R} , \lambda),\bar V_N  (G e ^{2 R} , \lambda)) $. Hence,  the pair $(V_X, X_P)$    
satisfies Eq.~\eqref{B3} as well as  Eq.~\eqref{SUR2}.
 This implies Eq.~\eqref{cororor}. 
 \hfill$\blacksquare$

In the proof of Corollary we have shown that any multi-parameter estimator $\{ m_i, X_i, P_i\}$ can be described by the measurement model $\{\mathcal E, \hat M, \hat N\}$.% with commutable $\hat M$ and $\hat N$. 
We can show that the converse direction is also true, and the two descriptions are equivalent as a type of Naimark's theorem.

\textbf{Theorem.---} 
For any joint measurement described by $\{\mathcal E, \hat M, \hat N\}$ with $[\hat M, \hat N]=0$, there exists a multi-parameter estimator $\{\hat m_i^\prime , a_i, b_i\}$ that gives the MSDs $(\bar V_M, \bar V_N )$.  The converse direction also holds. 

\textbf{Proof.---} 
Since $[\hat M, \hat N]=0$, there exists an orthonormal basis $\{\ket{\omega_i}\}$ that simultaneously diagonalizes $(\hat M,\hat N)$ such that $\hat M = \sum _i a_i \ketbra{\omega_i}{\omega_i} $  and  $\hat N = \sum _i b_i \ketbra{\omega_i}{\omega_i} $ with the sets of eigenvalues $\{a_i\}$ and  $\{b_i\}$.  Since $\mathcal E$ is a quantum channel, we have a Kraus representation $\mathcal E (\rho) = \sum_j K_j \rho K_j ^\dagger $ with $\sum_j K_j^\dagger  K_j  = \openone  $. 
Hence,  we can write 
\begin{align}
\bar V_M^{(x)} (\eta_M, \lambda) 
=& \sum_{i} \int p_\lambda (\alpha )(a_i - \sqrt{ \eta_M}  x_\alpha )^2  \tr (  \hat  m_i^\prime   \rho_\alpha ) d^2 \alpha , \nonumber \\
\bar V_N ^{(p)}(\eta_N, \lambda) 
= &\sum_{i} \int p_\lambda (\alpha )(b_i - \sqrt{ \eta_N}  p_\alpha )^2  \tr (  \hat  m_i^\prime   \rho_\alpha ) d^2 \alpha %\nonumber
\end{align}
where $\hat m_i^\prime :=  \sum_j  K_j \ketbra {\omega_i}{\omega_i}   K_j^\dagger \ge 0 $. From   $\sum_j K_j^\dagger  K_j  = \openone  $ and $ \sum_i \ketbra{\omega_i}{\omega_i} = \openone $   we can readily  check that $\{\hat m_i^\prime\}$ fulfills the condition for a POVM, 
  $\sum_i  \hat m_i^\prime   = \openone  $.  
This confirms that the pair $(\bar V_M, \bar V_N  )$ represents the MSEs in the form of  Eq.~\eqref{QESmse}. 
Therefore, the pair $(\bar V_M, \bar V_N)$ can be determined by the  property of an estimator $\{\hat m_i ^\prime, a_i, b_i\}$.  Conversely,  for any estimator $\{\hat m_i, X_i,P_i \}$ we can find a set $\{\mathcal E, \hat M, \hat N \}$ that gives the pair of the MSEs $(V_X, V_P)$ as shown in the proof of 
  Corollary.
    \hfill$\blacksquare$

Finally, we will reproduce one of the central outcomes in the quantum estimation theory.
For the mean-value estimation for  Gaussian states with the variances $(\sigma_x^2,\sigma_p^2)$, the most informative bound for  weighted MSEs \cite{holevo1975some} is given by 
 (see Eq.~(6.6.65) of Ref.~\cite{holevo2011probabilistic}) 
\begin{align}\label{MIB}
  g_x V_x +  g_p V_p \ge   g_x \sigma_x^2 +  g_p \sigma_p^2+ \sqrt{g_xg_p}.    \end{align}
We can immediately reach this relation from Eq.~\eqref{SUR2} for the case of pure Gaussian states, i.e.,  $\sigma_x^2 \sigma_p^2 = 1/4$, as follows. 
 Let us set $(\eta,\lambda) = (1,0)$ and take the square root of Eq.~\eqref{SUR2}. Then, by using the relation $(a t+  b t^{-1} )/2 \ge \sqrt{a b } $ for positive numbers $(a,b,t)$, we obtain  
\begin{align}\label{aaaaaa}
  t(\bar V_M(s,0) -  s/2) + t^{-1}(\bar V_N (s^{-1},0) -  s^{-1}/2)  \ge 1.    \end{align}
We can  see that
 Eq.~\eqref{aaaaaa}  coincides with  Eq.~\eqref{MIB}  by applying the   following replacement: $(\bar V_M(s,0), \bar V_M(s^{-1},0)) \to ( V_x ,V_p  )$,  $(s, s^{-1}) \to 2 (\sigma_x^2, \sigma_p^2) $, and $ t \to \sqrt{g_x / g_p  } $ as long as  $\sigma_x^2 \sigma_p^2 = 1/4$ holds. To this end, our approach reveals that the origin of the bound rather directly comes from commutation relations. The case for mixed Gaussian states should be addressed elsewhere. From a geometrical point of view,  Eq.~\eqref{MIB}  corresponds to tangent lines of the curve of  Eq.~\eqref{SUR2} (See Fig.~\ref{fig:smfig1.eps}).

We have presented a joint-measurement uncertainty relation based on a Bayesian input ensemble of optical states. It reproduces the form of the Arthurs-Kelly relation in Eq.~\eqref{measUR} in the uniform prior limit, and  the most informative bound for quantum estimation of phase-space displacement in the case of pure Gaussian states. Our measurement model is equivalent to the parameter estimation based on the POVM. Our uncertainty relation is applicable to such a general measurement process and, one can determine to what extend the performance of a given joint-measurement device is close to the theoretical limit in terms of the MSDs by using a moderate set of input states within realistic assumptions.

This work was supported by ImPACT Program of Council for Science, Technology and Innovation (Cabinet Office, Government of Japan).

%%%%%%%%%%%%%%%%%%%%%%%%%%%%%%%%%%%%%%%%%%%%%%%%%%%%%%%%%%%%%%%%%%%%%
%%%%%%%%%%%%%%%%%%%%%%%%%%%%%%%%%%%%%%%%%%%%%%%%%%%%%%%%%%%%%%%%%%%%%
%%%%%%%%%%%%%%%%%%%%%%%
%%%%%%%%%%%%%%%%%%%%%%%
\end{document}